# Slow Light Augmented Unbalanced Interferometry for Extreme Enhancement in Sensitivity of Measuring Frequency Shift in a Laser

**Ruoxi Zhu[1,*], Zifan Zhou[1], Jinyang Li[2], Jason Bonacum[3], David D. Smith[4], and Selim M. Shahriar[1,2,3]**
[1]*Department of Electrical and Computer Engineering, Northwestern University, Evanston, IL 60208, USA*
[2]*Department of Physics and Astronomy, Northwestern University, Evanston, IL 60208, USA*
[3]*Digital Optics Technologies, Rolling Meadows, IL 60008, USA*
[4]*NASA Marshall Space Flight Center*
[*]*Email: ruoxizhu2024@u.northwestern.edu*

**Abstract**

We demonstrate a slow-light augmented unbalanced Mach-Zehnder interferometer (MZI) which can be used to enhance very significantly the sensitivity of measuring the frequency shift in a laser. The factor of enhancement depends on the group index of the slow-light medium, the degree of imbalance between the physical lengths of the two arms of the MZI, and the spectral width of the laser. For a laser with a quantum noise limited spectral width, the group index has to be larger than the finesse of the laser cavity in order to achieve enhancement in measurement sensitivity. For the reported results, slow-light effect is produced by employing electro-magnetically induced transparency via coherent population trapping in a paraffin coated vapor cell of Rb atoms, with a maximum group index of ~2170. The maximum enhancement factor realized is ~70. This differs from the theoretically expected value of ~183 by a factor of ~2.6. This discrepancy can be attributed to effects of unidentified excess noise, and the fact that the test laser spectral width may not the quantum noise limited. Much larger values can potentially be obtained by modifying the apparatus, and using cold atoms for producing the slow-light effect. The sensitivity of any sensor that relies on measuring the frequency shift of a laser can be enhanced substantially using this technique. These include, but are not limited to, gyroscopes and accelerometers based on a conventional ring laser or a superluminal ring laser, and detectors for virialized ultra-light field dark matter.

## 1. Introduction

Many sensors rely on determining measurands based on shifts in frequencies of lasers. These include, but are not limited to, gyroscopes and accelerometers based on a conventional ring laser [1,2] or a superluminal ring laser [3], and detectors for virialized ultra-light field dark matter [4]. To achieve better precision in measuring frequency shifts, several schemes have been proposed employing superluminal lasers, which is a type of laser where the group velocity of light exceeds the speed of light in the vacuum [3,5,6,7,8,9,10,11,12,13,14,15,16]. However, it is also possible to achieve a sensitivity enhancement with a Slow-Light Augmented Unbalanced Mach-Zehnder Interferometer (SLAUMZI). The idea is closely related to the fact that the use of a slow-light medium inside an unbalanced Mach-Zehnder Interferometer (MZI) amplifies the number of fringes generated for a given change in the mean frequency, as shown by Boyd et al. [17]. However, their work did not take into account the spectral width of the laser, and how that affects the minimum measurable frequency shift (MMFS). Here, we determine the MMFS in a SLAUMZI, and how it depends on various parameters, including the finite linewidth of the laser and the degree of imbalance in the physical lengths of the two arms. Specifically, we identify the conditions under which the MMFS is smaller than what can be achieved using the standard technique [2,18] of measuring the beat frequency between the test laser and a reference laser with a spectral width that is negligibly small compared to that of the test laser. We find that for many scenarios of practical interest, the reduction in the MMFS can be very large. Furthermore, we present experimental results demonstrating such an enhancement.

## 2. Theoretical Model

The concept of slow-light augmented unbalanced interferometry (SLAUI) for the purpose of enhancing the sensitivity in measuring frequency shift can be realized using several different configurations. These include, for example, a Mach-Zehnder Interferometer (MZI) a Michelson Interferometer (MI), and a Fabry-Perot cavity (FPC). The mathematical analysis for the MZI and the MI are very similar. An analysis of the slow-light augmented FPC, which is inherently unbalanced, can be found in Ref. [19]. Experimental implementation of this concept is comparatively the least difficult for the MZI. As such, in this paper, we only present a study of the SLAUMZI, which is illustrated schematically in Figure 1.

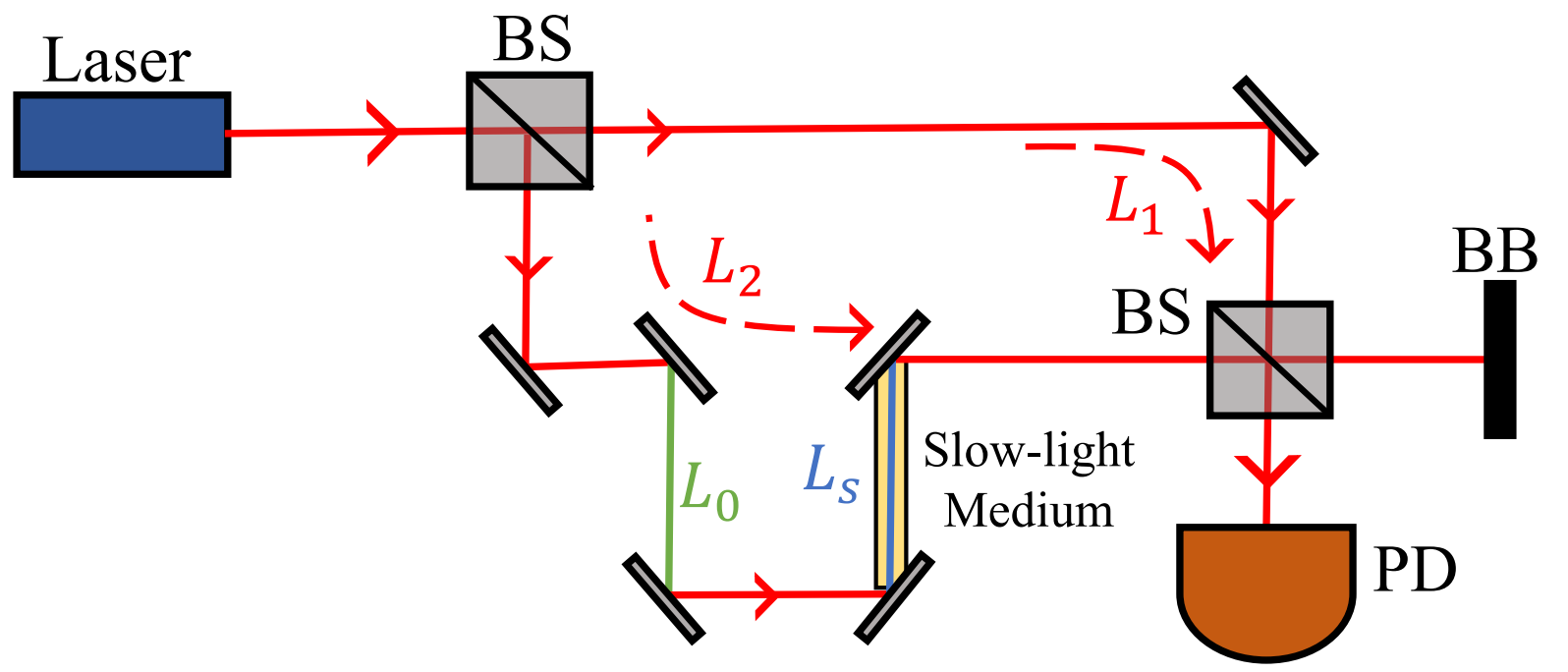

Figure 1. Illustration of the scheme for sensitivity-enhanced measurement of frequency shifts using an unbalanced Mach-Zehnder interferometer. The laser output is split by a beam splitter (BS) and the two beams propagate through different path length $L_1$ and $L_2$. A slow-light medium is placed inside one of the arms to generate the slow-light effect. This system produces a frequency dependent signal, observed by a photodetector in one of the output ports while the other output is blocked by a Beam Blocker (BB).

Here, we denote the length of the upper arm as $L_1$, and the length of the lower arm as $L_2$, which can be expressed as $L_2 = L_1 + L_0 + L_s$, where $L_s$ is the physical length of the slow-light medium and $L_0$ is an additional path-length difference. In what follows, we also assume the thickness of the windows of the glass cell containing the slow-light medium to be negligible. We denote the frequency as $\omega$, and the refractive index experienced by the laser field in the slow-light medium as $n(\omega)$. The phase difference between the two paths can be expressed as $\Phi(\omega) = L_0\omega/c_0 + n(\omega)L_s\omega/c_0$

, where $c_0$ is the vacuum speed of light. Therefore, the variation in the phase difference due to a change in the frequency can be expressed as $d\Phi/d\omega = L_0/c_0 + n_g L_s/c_0$, where $n_g = \partial(n\omega)/\partial\omega$ is the group index experienced by the laser at frequency $\omega$. We denote as the reference point the condition when the slow-light medium is replaced by vacuum, corresponding to unity group index (i.e., $n_g = 1$). In what follows, we will refer to the condition of $n_g > 1$ as the slow-light case, and the condition of $n_g = 1$ as the vacuum case. The fringe magnification factor, denoted as $M$, is then defined as the ratio between the phase difference variation in the slow-light case and the same for the vacuum case:

$$M \equiv \frac{(d\Phi/d\omega)_{SL}}{(d\Phi/d\omega)_{VAC}} = \frac{L_0/c_0 + n_g L_s/c_0}{L_0/c_0 + L_s/c_0} = \frac{\tau_0 + n_g \tau_D}{\tau_0 + \tau_D} \equiv \frac{\tau_{SL}}{\tau_{VAC}}, \qquad (1)$$

where $\tau_0 \equiv L_0/c_0$; $\tau_D \equiv L_S/c_0$; $\tau_{SL} \equiv \tau_0 + n_g \tau_D$; $\tau_{VAC} \equiv \tau_0 + \tau_D$ . We also define a parameter, $\rho \equiv \tau_D/(\tau_0 + \tau_D)$, which characterizes the effect of the additional path length difference, $L_0$, compared to the physical length of the slow-light medium, $L_S$. When there is no slow-light effect, we get $\tau_{SL} = \tau_{VAC}$ and $M = 1$. For $n_g \gg 1$, we see that $M \simeq \rho n_g$.

We assume the signal at the detection port is maximum when the phase difference between the two arms has a null value, and denote it as $S_0$. The signal as a function of the phase difference then can be expressed as $S_0 \cos^2[\Phi(\omega)/2]$. We define $\omega_0$ as the frequency at which the group index is evaluated, and denote $\Delta\omega$ as the deviation away from this frequency (i.e., $\Delta\omega \equiv \omega - \omega_0$). It is easy to see that the signal can be expressed as $S = S_0 \cos^2(\tau_{SL}\Delta\omega/2)$ by setting $\Phi(\omega_0)/2$ to zero without loss of generality.

This expression applies to the situation where the laser linewidth is assumed to be zero. Of course, in practice, the laser has a finite linewidth, greater than or equal to its quantum noise limited value known as the Schawlow-Townes Linewidth (STL). The STL can be expressed as follows [2,20,21,22,23]:

$$\gamma_{STL} = \frac{\hbar\omega\gamma_c^2}{2P_0} = \frac{\gamma_c^2}{2N} \equiv \frac{1}{\tau_{STL}}, \qquad (2)$$

where $P_0$ is the laser output power, $\gamma_c$ is the laser cavity linewidth, given by the inverse of the laser cavity decay time, $\tau_{LC}$, and $N$ is the number of output photons per second [24]. In this case, $\tau_{STL}$ represents the coherence time of the laser. The STL results from the loss of coherence due to spontaneous emission, and can be derived using many different ways, including the master equations approach, the Langevin noise operator approach, and the random walk approach [21,25], all yielding essentially the same result. As derived in Section A in the Supplement, the output of the SLAUMZI for an STL-limited laser can be expressed as:

$$S = \frac{S_0}{2} + \frac{S_0}{2}\exp\left(-\frac{\tau_{VAC}}{2\tau_{STL}}\right)\cos(\tau_{SL}\Delta\omega). \qquad (3)$$

For any sensor that depends on measuring the frequency shift of a laser, the key parameter is the minimum measurable frequency shift (MMFS). In Ref. [26], we have presented a detailed analysis of the MMFS of a laser, for different experimental modalities: an unbalanced MZI (UMZI), a Fabry-Perot Cavity (FPC), and heterodyning with a reference laser. We have shown that in all these cases the MMFS is given by the rms of the MMFS caused by phase-diffusion due to spontaneous emission, and the shot-noise caused by the vacuum mode that accompanies a coherent state. We have also shown that in all cases conditions exist for which the MMFS approaches what is known as the standard value of the MMFS: $\Delta\omega_{STD} = \sqrt{\gamma_M \gamma_{STL}} \equiv \Gamma_0$. In Section A of the Supplement, we have expanded this analysis to the case of the SLAUMZI. This analysis shows that it is possible to reach a value of the MMFS that is much smaller than the standard value of $\Gamma_0$. The Sensitivity Enhancement Factor (SEF), denoted as $R$, is defined as the ratio of the standard value of the MMFS and the MMFS of the SLAUMZI. The general expression for $R$, in the limit of $n_g \gg 1$, which holds for the experiment described below, is given in Eqn. (20) in the Supplement. In the limit where $\tau_{LC} \ll \tau_{VAC}$ which also holds for the experiment described below, the value of the SEF is predicted to be $R \simeq \rho n_g \sqrt{\sigma\eta}$, as shown in Eqn. (21) in the Supplement, where $\rho \equiv \tau_D / (\tau_0 + \tau_D)$ as defined above, $\sigma$ is the factor of signal attenuation caused by the slow-light medium, and $\eta$ is the quantum efficiency of the detector.

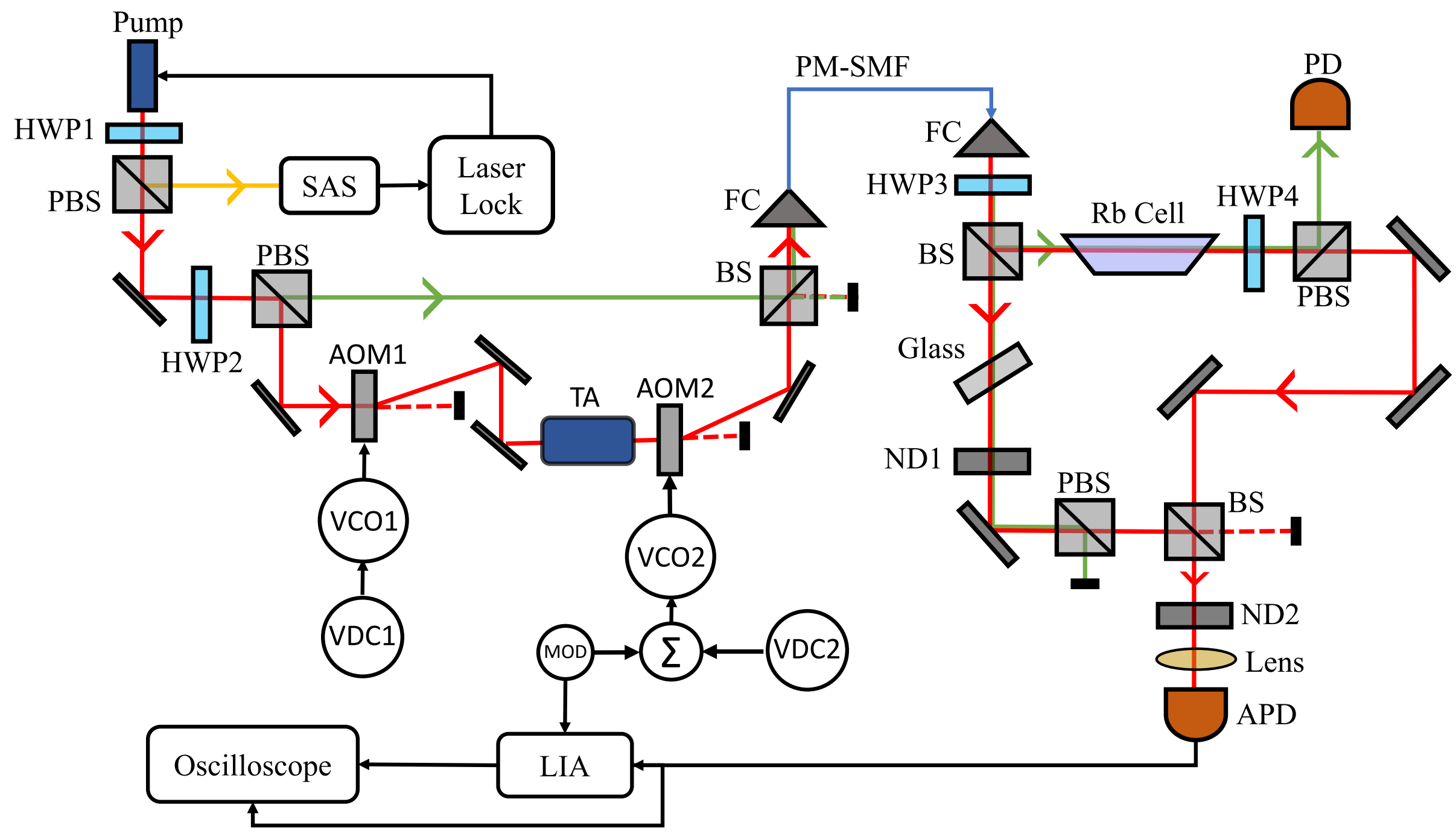

Figure 2. Experimental configuration for the EIT-augmented unbalanced MZI. The EIT process is generated by a pump and a probe that are two-photon resonant with the ground hyperfine states in $^{85}$Rb. The probe is produced from the pump using two acousto-optic modulators (AOMs) in a cascaded configuration. The pump and the probe, which are orthogonally polarized, are both sent into the MZI, and the pump is ejected before it reaches the output beam-splitter (BS). See **Methods** for additional details. APD: Avalanche Photo Diode; TA: Tapered Amplifier; ND: Neutral Density filter; SAS: Saturated Absorption Spectroscopy; HWP: Half Waveplate; PBS: Polarizing Beam-Splitter; FC: Fiber Coupler; VCO: Voltage Controlled Oscillator; VDC: DC voltage; MOD: Modulation; LIA: Lock-in Amplifier; PM-SMF: Polarization-Maintaining Single-Mode Fiber.

## 3. Results

Our primary objective is to demonstrate significant reduction in the MMFS using a SLAUMZI apparatus. We constructed a SLAUMZI in a manner that allows us to vary the group index from a few thousand down to unity, corresponding to an UMZI, as shown in Figure 2 and described in detail in **Section (a)** in **Methods**. This approach has made it possible for us to investigate the reduction in the MMFS for a SLAUMZI, in which Electromagnetic Induced Transparency (EIT) is utilized for generating the slow-light effect. EIT is a two-photon process that can produce a group index as large as ~$1.8*10^7$ [27]. It employs the use of two co-propagating laser beams passing through a Rb vapor cell, for example. Each laser sees a suppression of absorption when the two-photon resonance condition occurs.

### a. EIT measurement, fitting and characterization

Figure 3(a) illustrates schematically the atomic transitions used for generating EIT. The pump laser is locked to the transition coupling $5S_{1/2}\ F=3$ (state $|2\rangle$) to $5P_{1/2}\ F'=3$ (state $|4\rangle$). Maximum EIT occurs when the probe laser is tuned to the transition coupling $5S_{1/2}\ F=2$ (state $|1\rangle$) to state $|4\rangle$. Since the separation between the hyperfine states in the $5P_{1/2}$ manifold is less than the Doppler broadening in the vapor cell, $5P_{1/2}\ F'=2$ (state $|3\rangle$) also contributes significantly to the EIT process. Figure 3(b) shows the observed EIT signal, along with the theoretical fitting. The measured value of the fringe magnification factor, $M$, at the peak of the EIT signal, is ~380. Using Eq.(1), with $\tau_0 = 1.51$ nsec and $\tau_D = 0.27$ nsec, the corresponding group index is inferred to be ~2520, which in turn means that the value of $\tau_{SL}$ is ~674 nsec. We also note that the value of the parameter $\rho$ is ~0.15 for this case. The attenuation factor, $\sigma$, is ~0.63. The group index determined from the fitted result is ~2170. The difference between the experimental and theoretical values of the group index is reasonably small, and may be attributed to the fact that the theoretical model is somewhat approximate, since it does not take into account all the Zeeman sublevels explicitly. The processes of inferring the group index from $M$ and fitting the EIT signal are described in **Methods.** The probe laser has a linewidth of 376 kHz. This was determined by measuring the pair-wise beat spectra among three lasers. We assume that this linewidth is quantum noise limited, so that $\gamma_{STL} \simeq 2.4*10^6$ sec$^{-1}$ , and $\tau_{STL} \simeq 423$ nsec. For the avalanche photodiode (APD) used in the SLAUMZI output, the quantum efficiency, $\eta$, is ~0.5.

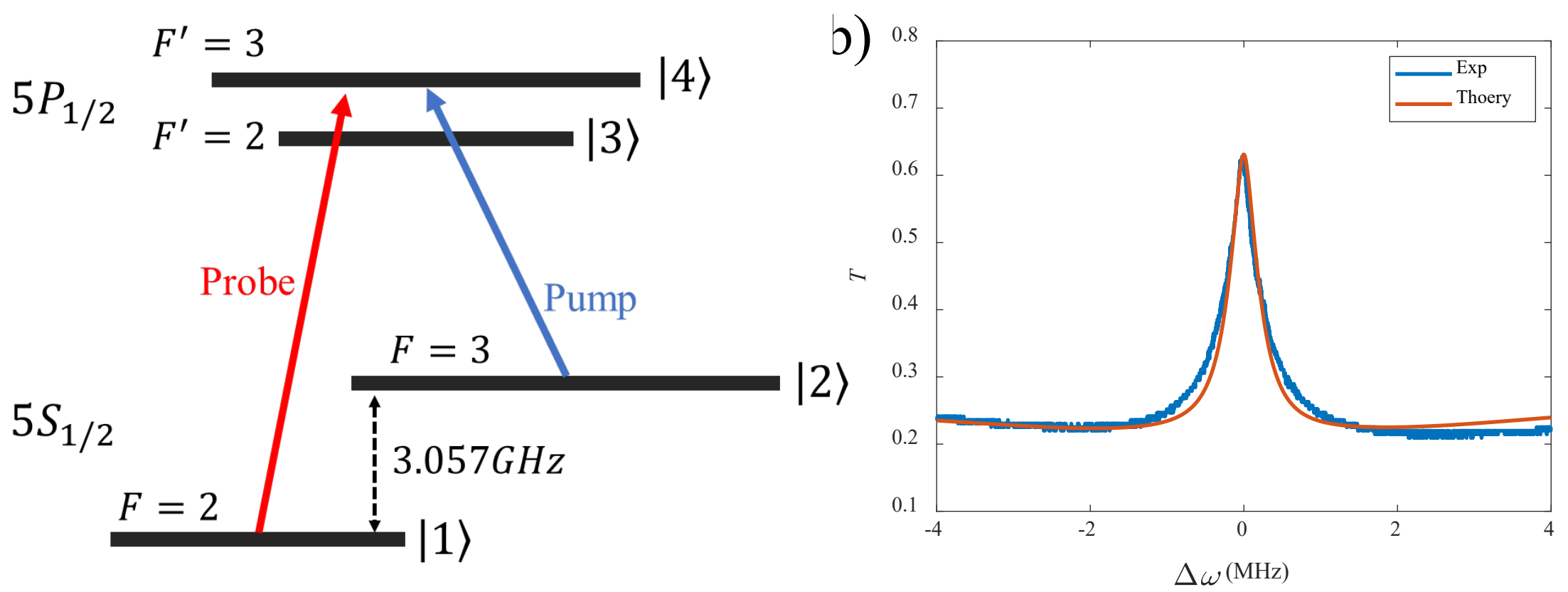

Figure 3. (a) Experimental configuration used for producing the EIT effect employing $^{85}$Rb. (b) The observed EIT signal, where the linewidth is ~0.5 MHz, and the fitted curve. The horizontal axis represents the frequency deviation from the center frequency $\omega_c = 3.0357\text{GHz}$ and the vertical axis denotes the normalized transmission.

To determine the value of the lifetime of the laser cavity, $\tau_{LC} \equiv 1/\gamma_c$, we note that it is related to the STL and the output power of the laser by the following expression $\tau_{LC} = \sqrt{\hbar\omega/2P_0\gamma_{STL}}$ , which follows from Eqn. (2). Given that the operating wavelength of the probe laser is ~795 nm, and the outpower power of the laser, $P_0$, is 21 mW, we find that the value of $\tau_{LC}$ is ~1.6*10$^{-12}$ sec. The value of $\tau_{VAC}$, which equals the sum of $\tau_0$ and $\tau_D$, is 1.78 nsec. As such, we find that $\tau_{LC}/\tau_{VAC}$ is ~8.9*10$^{-4}$.

The power incident at the input of the SLAUMZI was ~95 µW. However, in order to avoid saturating the avalanche photodiode (APD) used for detecting the SLAUMZI signal, the power was attenuated, using a neutral density filter, by a multiplicative factor of ~6.9*10$^{-3}$. Therefore, the effective power used for the SLAUMZI experiment, denoted as $P_S$, was ~0.65 µW. Therefore, the value of the parameter, µ, defined as the ratio of $P_S$ and $P_0$, is ~7*10$^{-3}$ . Thus, the condition $\left(\tau_{LC}/\tau_{VAC}\sqrt{\mu}\right) \ll 1$ is clearly satisfied, so that the effective value of the SEF, denoted as R, is given by Eqn. (21) of the Supplement: $R \simeq \rho n_g \sqrt{\sigma\eta}$ . Using $\rho \simeq 0.15$, $\sigma \simeq 0.63$, $\eta \simeq 0.5$, and $n_g \simeq 2170$, we find that the expected value of the SEF is $R \cong 183$ .

**b. MMFS measurement**

We measured the MMFS for the SLAUMZI by modulating the probe frequency and monitoring the demodulated signal. Specifically, the probe frequency was modulated around its quiescent value using a square wave applied to the VCO that drives AOM2. The glass plate was adjusted to maximize the signal modulation observed by the APD at the output port, thus ensuring operation at the point where the slope of the fringe is maximum. The APD signal was then fed into the LIA and demodulated after optimizing its phase. The swing in the laser frequency, denoted as $\omega_m$, due to the application of the square wave modulation, was varied while observing the LIA output.

Figure 4 shows the output of the LIA, denoted as $V_{LIA}$, as a function of the frequency swing, $\omega_m$. The value of $V_{LIA}$ is proportional to $\omega_m$ only when the modulated signal is significant compared to the excess noise. As $\omega_m$ decreases, the excess noise becomes dominant and the curve approaches a constant value asymptotically. To obtain the value of the MMFS, we assume the excess noise to have a constant value, and define the MMFS as the frequency at which the mean value of the signal equals the noise level. By implementing the fitting, we get the MMFS for the SLAUMZI system to be ~ 553 sec$^{-1}$.

The measurement bandwidth, $\gamma_M$, is given by the bandwidth of the Lock-in Amplifier (LIA), which is ~ 628 sec$^{-1}$, since it is slower than the APD, which has a bandwidth of ~ 3.14*10$^8$ sec$^{-1}$.

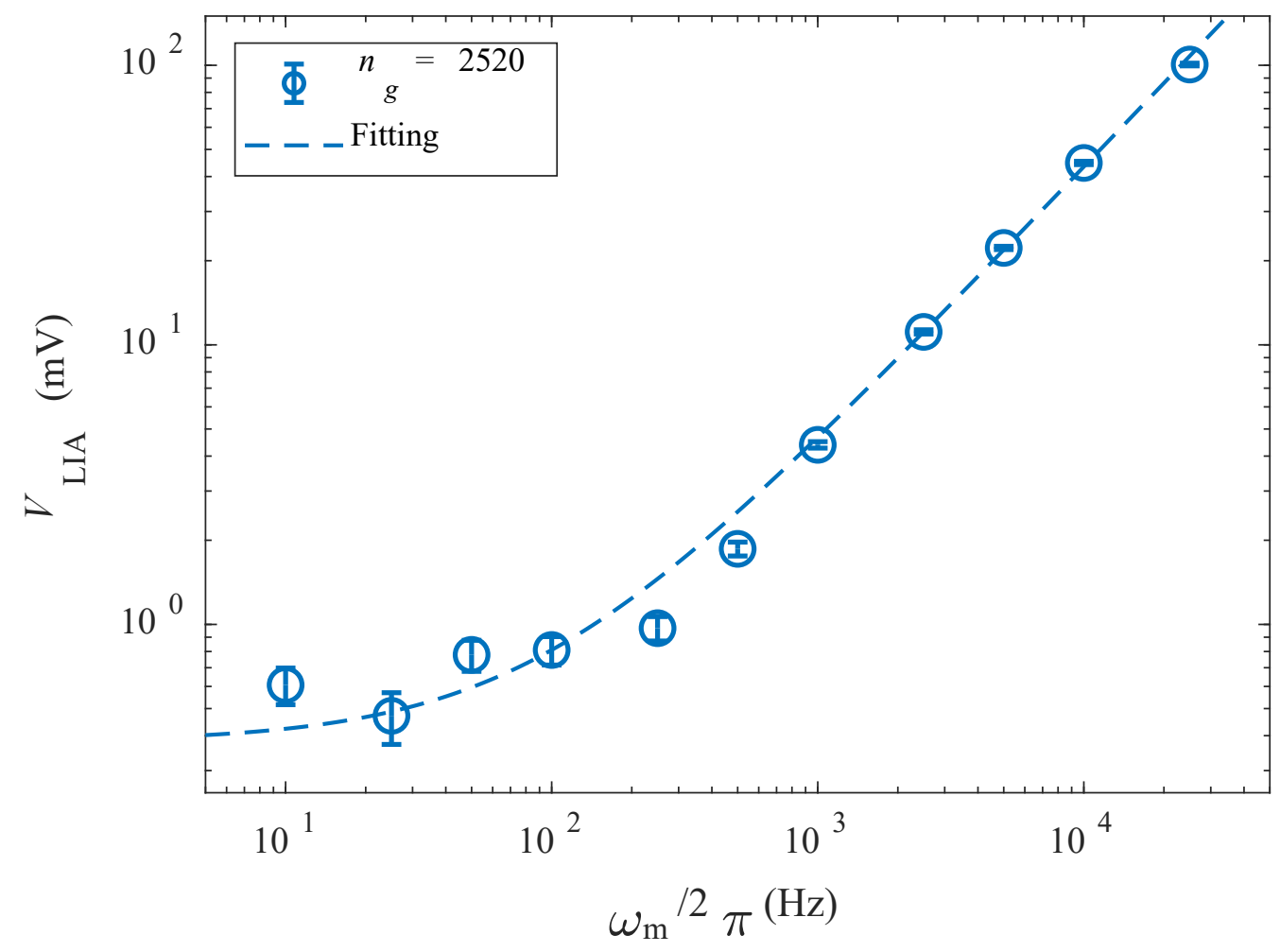

Figure 4. The output of the Lock-in amplifier as a function of the FM frequency deviation.

Given the value of $\gamma_{STL}$ mentioned above, the MMFS for the standard technique would be $\Delta\omega_{STD} \equiv \sqrt{\gamma_M \gamma_{STL}} = 3.85*10^3 \text{ sec}^{-1}$. The measure value of the SEF is therefore $R = 3.85*10^3 / 553 \simeq 70$.

This observed value of the SEF is smaller by a factor of ~2.6 than the theoretically expected value of ~183 mentioned above. The difference between the theoretical and experimental values can be attributed to several reasons. First, the excess noise present in the experiment is expected to limit the value of the SEF. A possible solution for suppressing it using balanced detection, by monitoring signals at both ports of the SLAUMZI apparatus. Second, it is difficult to determine whether the spectral width of the laser is indeed the STL, since determining the value of the STL requires independently measured value of the decay rate of the laser cavity, as well as the value of the Henry factor [21], which depends on the degree of the detuning between the lasing frequency and the peak of the gain spectrum of the lasing medium. Despite the disagreement with the ideal value, we believe that demonstration of such a large factor of enhancement in the value of the MMFS represents a major milestone in the field of precision sensing via measurement of frequency shifts.

It should also be noted that the value of the SEF would have been larger by a factor of $1/\rho \simeq 1/0.15 \simeq 6.67$ if the additional path length difference, $L_0$, were set to zero, thus making $\tau_0 = 0$, which would correspond to $\rho = 1$. However, for this initial experiment, we chose to use a substantial value of $L_0$ in order to make the fringes generated for the UMZI case to be clearly observable within the limitation of the scan range of the laser frequency. The UMZI fringes in turn were used to determine the fringe magnification factor, $M$, which enabled us to infer an estimated value of the group index simply from this factor, thus rendering confidence in the value of the group index inferred from fitting the observed EIT signal using a theoretical model. In the near future, we will implement a version of the experiment for which $\rho = 1$, so that even for this value of the group index, the theoretically expected SEF would be close to $\sim 183 * 6.67 \simeq 1221$ ~183*6.67. If the experimentally observed value of the SEF is again a factor of ~2.6 smaller, it would still reach a value of ~470. If course, actual value achievable experimentally would depend on the nature of excess noise.

## 4. Methods

### a. SLAUMZI Setup

Figure 2 shows the layout of the SLAUMZI, in which Electromagnetic Induced Transparency (EIT) is utilized for generating the slow-light effect. EIT is a two-photon process that can produce a group index as large as ~$1.8*10^7$ [27]. It employs the use of two co-propagating laser beams passing through a Rb vapor cell, for example. Each laser sees a greatly reduced absorption when the frequency difference between them matches the hyperfine splitting in the ground state. This induced transparency is due to the formation of a dark state. Here, the probe beam is generated by upshifting the frequency of the pump laser by two Acousto-optic modulators (AOMs) in series, each operating around ~1.5 GHz. The first AOM is driven by a voltage-controlled oscillator (VCO). The second AOM is driven by a frequency synthesizer. A tapered amplifier is utilized to increase the power of probe beam before the second AOM. The shifted frequency after the second AOM nominally matches the frequency difference (~3.0357 GHz) between the hyperfine states $5S_{1/2}\ F = 2$ and $5S_{1/2}\ F = 3$ in $^{85}$Rb, and can be tuned around this value. The frequency of the pump laser is locked to the $5S_{1/2}\ F = 3$ to $5P_{1/2}\ F' = 3$ transition via saturated absorption spectroscopy (SAS). The pump and the probe are combined with a 50/50 beam-splitter (BS). A polarization-maintaining single mode fiber (PM-SMF) is then used for mode matching of the pump and the probe, which are orthogonally polarized. The output of this fiber is first passed through a half-wave plate (HWP3), and then sent through another 50/50 BS at the input of the SLAUMZI.

The pump beam along the lower arm is ejected from the SLAUMZI by using a PBS. The fidelity of this ejection process is maximized by adjusting the angle of HWP3, to compensate for any residual rotation produced by the PM-SMF. Before measuring the MMFS, we rotate the halfwave plate HWP4 to eject the probe beam from the upper arm, and monitor it with a photodetector (PD), to make sure that the SLAUMZI is operating at the peak of the EIT signal. When measuring the MMFS, the orientation of HWP4 is adjusted to let the probe propagate but eject the pump beam from the upper arm. The angle of a glass plate inserted on the lower arm of the SLAUMZI can be rotated by a PZT-driven holder in order to adjust the bias phase difference between the paths. A neutral density filter (ND1) is placed in the lower arm after the glass plate in order to match, as closely as possible, the difference in the attenuation caused by the glass plate and sum of the residual attenuation caused by cell walls and the attenuation at the peak of the EIT signal. To minimize the effect of loss of coherence from collisions with the cell walls, we use a paraffin coated vapor cell. The windows are oriented at the Brewster's angle for the probe laser. The cell is placed inside a $\mu-$metal shield. The two arms of the SLAUMZI are combined with another 50/50 BS. The cell is ~9 cm long, and heated to a nominal temperature of ~37 degrees Celsius, as measured by a thermocouple gauge wrapped in a fiber glass sleeve. The signal is observed with an avalanche photo diode (APD) in one of the output ports of this BS. Another neutral density filter (ND2) is placed before the APD to prevent saturation. The output of the APD is monitored with an oscilloscope, and is also fed to a lock-in amplifier (LIA).

**b. Measurement of the Fringe Magnification Factor**

We start by calibrating the parameters of the SLAUMZI apparatus for the unity group index case. Specifically, we tune the probe frequency far away from any atomic resonance, and keep the atoms at room temperature. We also determine the approximate value of $\tau_{VAC}$ by measuring the values of $L_0$ and $L_s$ (see Figure 1). We then scan the probe frequency over a range of $\sim(2\pi/\tau_{VAC})$, producing at least one full spectral fringe. The spectral period of this fringe, defined as $\Delta\omega_{VAC}$ is measured carefully to determine a more accurate value of $\tau_{VAC}$, given by $(2\pi/\Delta\omega_{VAC})$.

To implement the EIT effect in the SLAUMZI, the probe beam is generated via upshifting the frequency of the pump beam by cascading two acousto-optic modulators (AOM1 and AOM2). The value of the net frequency shift is tuned to match the splitting between the hyperfine ground states. The slope of the SLAUMZI signal as a function of the probe frequency is expected to be maximum under this condition. If the group index remained constant over a range of probe frequencies larger than $(\Delta\omega_{VAC}/M)$, where $M$ is the fringe magnification factor, then it would be possible to observe a full spectral fringe by scanning the probe frequency, with a width of $\Delta\omega_{SL}=\Delta\omega_{VAC}/M$. However, the width of the EIT signal is typically narrower than this value, so that the group index has a constant (maximum) value over a range smaller than $\Delta\omega_{SL}$. We ensure that the measurement of M is carried out by staying within this limited range.

The value of $M$, which is used to infer the corresponding group index, is determined as follows. First, we scan the angle of the glass plate to determine the angles, denoted as $\theta_1$ and $\theta_2$, which produce the maximum ($V_{max}$) and minimum ($V_{min}$) values of the signal observed by the APD, respectively. We then tune the angle ~ halfway between these two values. Next, we apply a square wave at 10 kHz to the voltage controlled oscillator (VCO) that feeds AOM2. The amplitude of the square wave is chosen to produce a frequency swing of $\pm W$ for the probe beam. An iterative process is used to ensure that the value of $W$ is reasonably close to the value beyond which the signal varies non-linearly with $W$. For the data presented here, the value of $W$ used is 50 kHz. This square wave signal is also used as the external reference for the LIA. We then adjust the angle of the glass plate until the peak-to-valley amplitude of the modulation, defined as $\Delta V_{pp}$, observed at the APD output, is maximized, corresponding to operating at a point on the fringe with the maximum slope. We denote this value as $(\Delta V_{pp}^{\max})_{EIT}$. It should be noted that $(V_{\max}-V_{\min})/V_{\max}$ represents the visibility of the spectral fringe in the presence of the EIT effect.

In the absence of the EIT effect, which corresponds to the vacuum case, we can construct the expression for the spectral fringe as $V(\Delta\omega,\phi_0)=V_0\cos^2\left[(\tau_{VAC}\Delta\omega+\phi_0)/2\right]+V_{\min}$, where $V_0=(V_{\max}-V_{\min})$ and $\phi_0$ is an arbitrary phase offset. This ensures that the visibility of the spectral fringe in the vacuum case is the same as that for the EIT case. Therefore, the amplitude of the maximum modulation signal for the vacuum case can be calculated as $\left(\Delta V_{pp}^{\max}\right)_{VAC} = \left|V\left(W,\pi/2\right) - V\left(-W,\pi/2\right)\right| = V_0 \sin\left(\tau_{VAC} W\right)$ , for $\phi_0 = \pi/2$ . Then the fringe magnification factor is given by:

$$M = \left| \frac{\left(\Delta V_{pp}^{\max}\right)_{EIT}}{\left(\Delta V_{pp}^{\max}\right)_{VAC}} \right| \tag{4}$$

**c. EIT measurement and modeling**

The EIT effect is modeled using the four-level system shown in Figure 3(a), by solving the Liouville equations [28]. In order to estimate the Rabi frequencies as well as the branching ratios of the decay rates, it is necessary to determine the effective dipole transition matrix elements for each of the four hyperfine transitions ($|1\rangle \rightarrow |3\rangle$, $|1\rangle \rightarrow |4\rangle$, $|2\rangle \rightarrow |3\rangle$, and $|2\rangle \rightarrow |4\rangle$). We do this by using a weighted averages of the dipole transition matrix elements among the Zeeman sublevels, for linear polarizations (sum of left and right circular polarizations), multiplied by a fitting parameter, $\alpha$, that is different for each transition. In Figure 3(b), the blue trace shows the experimental observed EIT signal, and the red trace shows the theoretical result, obtained for the following values of the fitting parameters: $\alpha_{13} = 0.37, \alpha_{14} = 0.20, \alpha_{23} = 0.24, \alpha_{24} = 1.18$. The corresponding probe Rabi frequencies for the 1-3 and 1-4 transitions are $\Omega_{13} = 8.3*10^6$ sec$^{-1}$ and $\Omega_{14} = 1.1*10^7$ sec$^{-1}$, and the pump Rabi frequencies for the 2-3 and 2-4 transitions are $\Omega_{23} = 1.6*10^7$ sec$^{-1}$ and $\Omega_{24} = 1.0*10^7$ sec$^{-1}$, respectively, determined using experimentally measured intensities. The decay rates for the 3-1 and 3-2 transition are $2.6*10^7$ sec$^{-1}$ and $5.2*10^6$ sec$^{-1}$, respectively. And the decay rates for the 4-1 and 4-2 transition are $1.0*10^7$ sec$^{-1}$and $3.1*10^7$ sec$^{-1}$, respectively. The collisional decay rates between the two ground states $|1\rangle$ and $|2\rangle$ are assumed to be symmetric, and treated as another fitting parameter. The best fitting obtained corresponded to a value of $3.8*10^5$ sec$^{-1}$ for this rate. As noted earlier, the Rb cell temperature was ~37 degrees Celsius, as measured by a thermocouple gauge encased in a fiber-glass sleeve. This may not represent the actual temperature of the atoms inside the cell. As such, we treated the temperature as another fitting parameter. The best fitting was obtained for a temperature of 29.6 degrees Celsius. The Doppler broadening is also accounted for by averaging the signal over different velocity groups. Moreover, since the absorption is significant, we implement the cell slicing technique in our model, in which we slice the Rb cell into 10 segments and calculate the intensity of the incident light after each slice.

**d. MMFS measurement and fitting**

We express the ideal value of the demodulated output of the lock-in amplifier (LIA) as $V_{ideal} = k\omega_m/(2\pi)$, where $\omega_m$ is the frequency deviation, and $k$ is a coefficient which can be obtained by fitting the linear part of the experimental data. The effect of excess noise is modeled as a constant parameter in the LIA output, the magnitude of which can be adjusted to give the best fitting result. If we denote the noise as $V_N$, the total output of the LIA can be written as $V_{fit} = k\omega_m/(2\pi) + V_N$ . The difference between the fitting value, $V_{fit}$, and the experimental value, $V_{LIA}$, can be defined as:

$$\Delta V_{diff} \equiv V_{LIA} - V_{fit} = V_{LIA} - \left( k\frac{\omega_m}{2\pi} + V_N \right) \tag{5}$$

We further define the sum of the squares of the ratio between the difference and the LIA output as $K_{sum}$:

$$K_{sum} = \sum_m \left[ \Delta V_{diff} \big/ V_{LIA} \right]^2 = \sum_m \left\{ \left[ V_{LIA} - k\frac{\omega_m}{2\pi} - V_N \right] \Big/ V_{LIA} \right\}^2 \tag{6}$$

By minimizing the value of $K_{sum}$, we determine the value of $V_N$ to be ~0.38 mV. Figure 4 shows the best fitting curve found by implementing the process described above. The linear part of this fit yields the value of $k$ to be ~4.3*$10^{-3}$ mV/sec. We adopt the criterion that the MMFS corresponds to the frequency for which the ideal signal equals the noise level: $V_{ideal} = V_N$. It then follows that the value of the MMFS is $2\pi V_N / k$.

---

[1]E. J. Post, "The Sagnac Effect," Rev. Mod. Phys. 39, 475 (1967).
[2]W. W. Chow, J. Gea-Banacloche, L. M. Pedrotti, V. E. Sanders, W. Schleich, and M. O. Scully, "The ring laser gyro," Reviews of Modern Physics 57, 61 (1987).
[3] M. S. Shahriar, G. S. Pati, R. Tripathi, V. Gopal, and M. Messall, "Ultrahigh enhancement in absolute and relative rotation sensing using fast and slow light," Phys. Rev. A 75, 053807-1 - 053807-10 (2007).
[4]A. A. Geraci, C. Bradley, D. Gao, J. Weinstein, and A. Derevianko, "Searching for Ultralight Dark Matter with Optical Cavities," Phys. Rev. Lett. 123, 031304 (2019).
[5] H. N. Yum, M. Salit, J. Yablon, K. Salit, Y. Wang, and M. S. Shahriar, "Superluminal ring laser for hypersensitive sensing," Opt. Express 18(17), 17658 (2010).
[6] J. Scheuer and S.M. Shahriar, "Lasing dynamics of super and sub luminal lasers", Opt. Express 23, 32350– 32365 (2015).
[7] Y. Wang, Z. Zhou, J. Yablon, and M. S. Shahriar, "Effect of Multi-Order Harmonics in a Double-Raman Pumped Gain Medium for a Superluminal Laser," Opt. Eng. 54(5), 057106 (2015).
[8] J. Yablon, Z. Zhou, M. Zhou, Y. Wang, S. Tseng, and S.M. Shahriar, "Theoretical modeling and experimental demonstration of Raman probe induced spectral dip for realizing a superluminal laser," Optics Express, Vol. 24, No. 24, 27446 (2016).
[9] M. Zhou, Z. Zhou, M. Fouda, N.J. Condon, J. Scheuer, and M.S. Shahriar, "Fast-light Enhanced Brillouin Laser Based Active Fiber Optic Sensor for Simultaneous Measurement of Rotation and Strain," Journal of Lightwave Technology, 35 (23), 5222 (2017).
[10] M.F. Fouda, M. Zhou, H.N. Yum, and S.M. Shahriar, "Effect of cascaded Brillouin lasing due to resonant pumps in a superluminal fiber ring laser gyroscope," Opt. Eng. 57(10), 107108 (2018).
[11] Z. Zhou, M. Zhou and S.M. Shahriar, "A superluminal Raman laser with enhanced cavity length sensitivity," Optics Express 27, 29739-29745 (2019).
[12] Y. Sternfeld, Z. Zhou, J. Scheuer, and M. S. Shahriar, "Electromagnetically induced transparency in Raman gain for realizing a superluminal ring laser," Opt. Express 29, 1125-1139 (2021).
[13] Zifan Zhou, Nicholas Condon, Devin Hileman, and M. S. Shahriar, "Observation of a highly superluminal laser employing optically pumped Raman gain and depletion," Opt. Express 30, 6746-6754 (2022).
[14] Z. Zhou, R. Zhu, N. Condon, D. Hileman, J. Bonacum and S.M. Shahriar, "Bi-directional Superluminal Ring Lasers without Cross-talk and Gain Competition," Appl. Phys. Lett. 120, 251105 (2022).
[15] Z. Zhou, R. Zhu, Y. Sternfeld, J. Scheuer, J. Bonacum, and S. M. Shahriar, "Demonstration of a Superluminal Laser using Electromagnetically Induced Transparency in Raman Gain," Opt. Express 31(9), 14377-14388 (2023).
[16] Y. Sternfeld, Z. Zhou, S.M. Shahriar and J. Scheuer, "Single-pumped gain profile for a superluminal ring laser," Optics Express 31(22) 36952-36965 (2023).

[17] Z. Shi, Robert, W. Boyd, Daniel J. Gauthier, and C. C. Dudley, "Enhancing the spectral sensitivity of interferometers using slow-light media," Opt. Lett. 32, 915-917 (2007).
[18] F. Aronowitz, in Laser Applications, edited by M. Ross (Academic, New York) pp. 113-200.
[19]R. Zhu, Z. Zhou, D. Greenwood, J. Bonacum, D. D. Smith, and S. M. Shahriar, "Slow Light Augmented Fabry-Perot Cavity for Enhanced Sensitivity in Measuring Frequency Shift," https://doi.org/10.48550/arXiv.2506.15885
[20] C. H. Townes, "Some applications of optical and infrared masers," in *Advances in Quantum Electronics*, J.R. Singer., Ed., New York: Columbia Univ. Press, 1961, pp 1-11.
[21] C. H. Henry, "Theory of the Linewidth of Semiconductor Lasers," IEEE J. Of Quantum Electronics, Vol. QE-18, No. 2, (1982).
[22] T.A. Dorschner, H.A. Haus, M. Holz, I.W. Smith, H. Statz, "Laser gyro at quantum limit", IEEE J. Quant. Elect. QE-16, 1376 (1980).
[23] B. T. King, "Application of superresolution techniques to RLGs: exploring the quantum limit," Applied Optics 39, 6151 (2000).
[24] Here, we have assumed that the laser operates at the peak of the gain spectrum, so that the so-called Henry factor [21] can be ignored. The analysis that follows can be easily extended to the more general case where the Henry factor is dominant.
[25] M. O. Scully and M.S. Zubairy, Quantum Optics, Cambridge University Press (1997).
[26] S. M. Shahriar, R. Zhu, J. Li and Z. Zhou, "The General Quantum Limit for and the Optimization of the Minimum Measurable Frequency Shift in a Laser," https://arxiv.org/pdf/2607.10518
[27] L. V. Hau, S. E. Harris, Z. Dutton and C. H. Behrooz, "Light speed reduction to 17 metres per second in an ultracold atomic gas," Nature 397, 18 (1999).
[28] M. S. Shahriar, Y. Wang, S. Krishnamurthy, Y. Tu, G. S. Pati and S. Tseng, "Evolution of an N-level system via automated vectorization of the Liouville equations and application to optically controlled polarization rotation," Journal of Modern Optics, 61:4, 351-367 (2014).

# Supplement

Here we provide theoretical discussions that are relevant to the main body of the paper. The supplement is organized as follows. In Section A, we present an analysis of the MMFS for the SLAUMZI and comparison with the standard MMFS. In Section B, we present an explicit example of how the SLAUMZI technique can be applied to realize an ultra-sensitive gyroscope and accelerometer.

### Section A. Analysis of the MMFS for the SLAUMZI and Comparison with the Standard MMFS

In Ref. [1], we have shown the general limit for the MMFS of a laser for techniques that do not employ the slow-light medium. Specifically, we have shown that the MMFS is determined by two effects: phase diffusion caused by spontaneous emission, and shot-noise due to the vacuum mode that accompanies a coherent state. We also determined the MMFS for three different sensing modalities: an unbalanced MZI (UMZI), a Fabry-Perot Cavity (FPC), and heterodyning with a reference laser. In all cases, we showed that conditions exist for which the MMFS approaches the standard value:

$$\Delta\omega_{STD} = \sqrt{\gamma_M \gamma_{STL}} \equiv \Gamma_0 , \tag{1}$$

which is the geometric mean of the measurement bandwidth and the STL, while under some conditions the MMFS can be much larger than $\Gamma_0$. Here, we determine the MMFS for a SLAUMZI, and compare it with the standard MMFS.

To determine the MMFS for a SLAUMZI, we consider first the random phase jumps caused by spontaneous emission events in the laser. The probability of such jumps modeled as random walks follows a Gaussian distribution. It is well-known that such a model yields the proper expression for the STL [1,2]. To illustrate the effect of these phase jumps on the SLAUMZI signal, it is instructive to discuss first the vacuum case corresponding to unity group index. We define as $\tau_i$ the time for light to travel from the laser to the input port of the MZI, and $\tau_1 \left(\tau_2\right)$ as the time for light to travel along the upper (lower) arm of the MZI. Using the definitions summarized below Eqn. (1) in the main body, we then get $\tau_1 = L_1/c_0$ and $\tau_2 = L_2/c_0 = \tau_1 + \tau_0 + \tau_D = \tau_1 + \tau_{VAC}$. Thus, at the output port of the SLAUMZI, the field arriving via the upper arm at time $t$ is shifted in phase by the phase-jump, $\phi_1$, that occurred at a time $t-\left(\tau_i + \tau_1\right)$. On the other hand, the field arriving via the lower arm at time $t$ is shifted in phase by the phase-jump, $\phi_2$, that occurred at a time $t-\left(\tau_i + \tau_2\right)$. Therefore, the unity-group-index SLAUMZI signal can be expressed as:

$$\begin{aligned} S &= S_0 \cos^2\left[(\tau_{VAC}\Delta\omega + \phi_2 - \phi_1)/2\right] \\ &= \frac{S_0}{2} + \frac{S_0}{2}[\cos(\tau_{VAC}\Delta\omega)\cos(\Delta\phi) - \sin(\tau_{VAC}\Delta\omega)\sin(\Delta\phi)] \end{aligned} \tag{2}$$

where $\Delta\phi = \phi_2 - \phi_1$. Thus, the two instances when the laser phase jumps to $\phi_1$ and $\phi_2$ are separated by the time difference $\tau_{VAC} = \tau_2 - \tau_1$.

Consider next the slow-light case, corresponding to the non-unity group index. We define $\tilde{\tau}_1(\tilde{\tau}_2)$ as the time for light to travel along the upper (lower) arm of the MZI as $\tilde{\tau}_1 = L_1/c_0 = \tau_1$ and $\tilde{\tau}_2 = \tilde{\tau}_1 + \tau_0 + nL_s/c_0 = \tilde{\tau}_1 + \tau_0 + n\tau_D = \tilde{\tau}_1 + \tau_{ND}$, where $n$ is the phase index (*not* the group index) of the slow-light medium, and we have defined $\tau_{ND} \equiv \tau_0 + n\tau_D$. Since the experiment is carried out for a continuous wave (CW) laser, it is proper to use the phase index for the propagation of the mean field. If the spectral width of the laser is due strictly to the STL caused by spontaneous emission, the phase jumps can be modeled as occurring instantaneously (according to the Markov approximation used in modeling the coupling between the atoms and the bath of vacuum modes). On the other hand, the bandwidth of the positive dispersion that causes the slow-light effect is typically very small (of the order of a few MHz in the experiment reported here). As such, it is a good approximation to assume that the phase jump would also propagate through the slow-light medium at essentially the phase velocity, and not the group velocity. Without loss of generality, we see that $\tilde{\tau}_2 = \tilde{\tau}_1 + \tau_{ND}$. As such, the non-unity group index SLAUMZI signal can be expressed as:

$$\begin{aligned} S &= S_0 \cos^2\left[(\tau_{SL}\Delta\omega + \phi_2 - \phi_1)/2\right] \\ &= \frac{S_0}{2} + \frac{S_0}{2}[\cos(\tau_{SL}\Delta\omega)\cos(\Delta\phi) - \sin(\tau_{SL}\Delta\omega)\sin(\Delta\phi)] \end{aligned} \tag{3}$$

where $\Delta\phi = \phi_2 - \phi_1$, and the two instances when the laser phase jumps to $\phi_1$ and $\phi_2$ are separated by the time difference $\tau_{ND} = \tilde{\tau}_2 - \tilde{\tau}_1$. It should be noted that in this case the time parameter that determines the scale of the signal fringes is now $\tau_{SL} \equiv \tau_0 + n_g\tau_D$, which is much larger than $\tau_{ND} = \tau_0 + n\tau_D$ when the group index is substantially larger than unity, since the phase index is very close to unity if diluted vapor gas at close to room temperature or cold atoms are used for generating the slow-light effect.

Using a random walk model, the probability of having two instances of phase jumps separated by $\tau_{ND}$ can be expressed as [1,2]:

$$f(\Delta\phi_{\tau_{ND}}) = \frac{1}{\sqrt{2\pi\,\tau_{ND}/\tau_{STL}}}\exp\left(-\frac{\Delta\phi_{\tau_{ND}}^2}{2(\tau_{ND}/\tau_{STL})}\right). \tag{4}$$

Therefore, the output signal can be expressed as:

$$S = \frac{S_0}{2} + \frac{S_0}{2}\left[\cos(\tau_{SL}\Delta\omega)\left\langle\cos(\Delta\phi_{\tau_{ND}})\right\rangle - \sin(\tau_{SL}\Delta\omega)\left\langle\sin(\Delta\phi_{\tau_{ND}})\right\rangle\right]. \tag{5}$$

The expectation value $\left\langle\cos\left(\Delta\phi_{\tau_{ND}}\right)\right\rangle$ computed over the distribution shown in Eqn. (4) is found to be $\exp(-\tau_{ND}/2\tau_{STL})$, and it is obvious that $\left\langle\sin\left(\Delta\phi_{\tau_{ND}}\right)\right\rangle = 0$ because of the symmetry of $f(\Delta\phi_{\tau_{ND}})$ around the origin. Therefore, the output signal can be expressed as:

$$S = \frac{S_0}{2} + \frac{S_0}{2}\exp(-\frac{\tau_{ND}}{2\tau_{STL}})\cos(\tau_{SL}\Delta\omega)\,. \tag{6}$$

The phase index can be approximated to be unity if diluted vapor gas at close to room temperature or cold atoms are used for generating the slow-light effect. We then get $\tau_{ND} = \tau_{VAC}$, so that the output signal for the SLAUMZI can be expressed as:

$$S = \frac{S_0}{2} + \frac{S_0}{2}\exp(-\frac{\tau_{VAC}}{2\tau_{STL}})\cos(\tau_{SL}\Delta\omega)\,. \tag{7}$$

For the case where no slow light is used, the expression for the signal becomes $S = (S_0/2)\cdot(1+\exp(-\tau_{ND}/2\tau_{STL})\cos(\tau_{ND}\Delta\omega))$. For the vacuum case, we have $n=1$, and the corresponding expression for the signal is:

$$S = \frac{S_0}{2} + \frac{S_0}{2}\exp(-\frac{\tau_{VAC}}{2\tau_{STL}})\cos(\tau_{VAC}\Delta\omega)\,. \tag{8}$$

This is consistent with the result that the interference signal disappears when the time delay greatly exceeds the coherence time. It should be noted that this expression for the signal for a conventional unbalanced MZI (UMZI) is well accepted. For example, the approach used in Ref. [3] for deriving the spectral width of a Fabry-Perot cavity can be easily adopted to get this expression for the UMZI.

The expression shown in Eqn. (7) can now be used to determine the MMFS for the SLAUMZI, using the approach developed in Ref. [1]. Specifically, the MMFS is given by

$$\Delta\omega_{MIN,SLAUMZI} = \sqrt{\Delta\omega_{MIN,SE}^2 + \Delta\omega_{MIN,VM}^2} \tag{9}$$

where $\Delta\omega_{MIN,SE}$ is the MMFS due to solely to phase diffusion caused by spontaneous emission, and $\Delta\omega_{MIN,VM}$ is the MMFS due to shot-noise caused by the vacuum mode. To determine the value of $\Delta\omega_{MIN,SE}$, we note first that since the phase index is assumed to be approximately unity, the probability distribution for the phase difference $\Delta\phi$ between the two arms of the SLAUMZI can be found from Eqn. (7) by using $\tau_{ND} = \tau_{VAC}$:

$$f(\Delta\phi) = \frac{1}{\sqrt{2\pi\,\tau_{VAC}/\tau_{STL}}}\exp(-\frac{\Delta\phi^2}{2(\tau_{VAC}/\tau_{STL})})\,. \tag{10}$$

The standard deviation of phase difference $\Delta\phi$ can be determined immediately from Eqn. (10) to be $\Delta\tilde{\phi} = \sqrt{\tau_{VAC}/\tau_{STL}}$. It then follows that:

$$\Delta\omega_{MIN,SE} = \frac{\Delta\tilde{\phi}}{\tau_{SL}} = \left(\frac{\tau_{VAC}}{\tau_{SL}}\right)\sqrt{\frac{1}{\tau_{VAC}\tau_{STL}}} = \frac{1}{M}\sqrt{\frac{1}{\tau_{VAC}\tau_{STL}}} \simeq \frac{1}{\rho n_g}\sqrt{\frac{1}{\tau_{VAC}\tau_{STL}}}\,. \tag{11}$$

Here, $M$ is the fringe magnification factor, and we have made the approximation that for $n_g >> 1$, $M \simeq \rho n_g$, where $\rho \equiv \tau_D / (\tau_0 + \tau_D)$, as defined in the text after Eqn. (1) in the main body. If the measurement is repeated $\nu = (\tau_M / \tau_{VAC})$ times, where $\tau_M = 1/\gamma_M$ is the measurement time, we get:

$$\Delta\omega_{MIN,SE} \simeq \frac{1}{\rho n_g}\sqrt{\frac{1}{\nu \tau_{VAC} \tau_{STL}}} = \frac{1}{\rho n_g}\sqrt{\frac{1}{\tau_M \tau_{STL}}} = \frac{1}{\rho n_g}\sqrt{\gamma_M \gamma_{STL}} = \frac{\Delta\omega_{STD}}{\rho n_g} = \frac{\Gamma_0}{\rho n_g}. \qquad (12)$$

We introduce the parameter, $\eta$, which accounts for the finite quantum efficiency of the detector, and the parameter, $\sigma$, which accounts for the effect of attenuation caused by the slow-light medium. When these effects are taken into account, we get:

$$\Delta\omega_{MIN,SE} \simeq \frac{\Delta\omega_{STD,EFF}}{\rho n_g \sqrt{\sigma\eta}} = \frac{\Gamma_0}{\rho n_g \sqrt{\sigma\eta}}. \qquad (13)$$

To determine the value of $\Delta\omega_{MIN,VM}$, we consider the variance of the signal due to shot-noise:

$$\Delta\omega_{MIN,VM} = \left|\frac{\Delta S_{SN}}{\partial S / \partial \Delta\omega}\right| = \left|\frac{\sqrt{S}}{\partial S / \partial \Delta\omega}\right|. \qquad (14)$$

where $\Delta S_{SN} = \sqrt{S}$ is the standard deviation of the signal due to shot-noise. For a typical experiment of interest, it is reasonable to assume that $\tau_{VAC} \ll \tau_{STL}$, so that for evaluating $\Delta\omega_{MIN,VM}$ we can write the signal for the SLAUMZI as:

$$S \simeq (S_0/2)(1+\cos(\tau_{SL}\Delta\omega)) \simeq S_0 \cos^2(M\tau_{VAC}\Delta\omega) = S_0 \cos^2(\rho n_g \tau_{VAC}\Delta\omega). \qquad (15)$$

where we have again made the approximation that for $n_g >> 1$, $M \simeq \rho n_g$. To evaluate this explicitly, we note first that the ideal method for measuring any shift in the frequency of the laser would make use of the so-called hopping process [4], which would work as follows for the SLAUMZI. The frequency of the laser would be modulated with a square wave, thus making it jump by a value of $\Delta\omega = \pm\left[\pi / (2\rho n_g \tau_{VAC})\right]$, and the signal observed at these two extrema (corresponding to maximum slopes) would be subtracted from one another. If the unmodulated value of the laser frequency corresponds to the peak of the signal, for example, then the subtracted signal would be zero, in the absence of any noise. Any deviation of the subtracted signal away from zero would then represent a shift in the laser frequency. It can be shown that the minimum frequency uncertainty obtained using this hopping method would be smaller by a factor of $\sqrt{2}$ to what one would obtain if the signal is measured at the point where it has the highest slope, corresponding to a bias frequency shift of $\Delta\omega = \pi / \left(2\rho n_g \tau_{VAC}\right)$. We determine the MMFS in both cases, by including a parameter, $\beta$, which has different values for the hopping method and the one-sided method. Specifically, we get:

$$\Delta\omega_{MIN,VM} = \frac{\beta}{\rho n_g \tau_{VAC}} \sqrt{\frac{1}{S_o}} . \qquad (16)$$

where $\beta = \sqrt{2}$ for the one-sided method, and $\beta = 1$ for the hopping method.

We introduce a parameter, $\mu = P_s / P_o$, where $P_s$ is the power incident on the SLAUMZI, and $P_o$ is the total output power of the laser. Taking into account the effects of the parameters $\mu$, $\eta$ and $\sigma$ introduced above, we can thus write:

$$S_o = \sigma\eta\mu N\tau_M = \sigma\eta\mu \frac{\langle n\rangle \gamma_{LC}}{\gamma_M} = \frac{\sigma\eta\mu\gamma_{LC}^2}{2\gamma_M\gamma_{STL}}, \qquad (17)$$

where $\langle n\rangle$ is the mean number of photons in the laser cavity. Inserting Eqn. (17) in Eqn. (16), we get:

$$\Delta\omega_{MIN,VM} = \frac{\beta\sqrt{2}}{\rho n_g} \frac{\tau_{LC}}{\tau_{VAC}} \sqrt{\frac{\gamma_M \gamma_{STL}}{\sigma\eta\mu}} = \beta\sqrt{2} \frac{\tau_{LC}}{\tau_{VAC}} \frac{\Gamma_0}{\rho n_g \sqrt{\sigma\eta\mu}} = \frac{\beta\sqrt{2}}{\sqrt{\mu}} \frac{\tau_{LC}}{\tau_{VAC}} \Delta\omega_{MIN,SE,} . \qquad (18)$$

The total MMFS is then found to be:

$$\Delta\omega_{MIN,SLAUMZI} = \Delta\omega_{MIN,SE} \sqrt{1 + \frac{2\beta^2}{\mu} \left( \frac{\tau_{LC}}{\tau_{VAC}} \right)^2} . \qquad (19)$$

Comparing Eqn. (19) with Eqn. (1), we find that the Sensitivity Enhancement Factor (SEF), denoted as R, is given by:

$$R = \frac{\Delta\omega_{MIN,STD}}{\Delta\omega_{MIN,SLAUMZI}} = \sqrt{\sigma\eta}\rho n_g \frac{1}{\sqrt{1 + \frac{2\beta^2}{\mu} \left( \frac{\tau_{LC}}{\tau_{VAC}} \right)^2}} . \qquad (20)$$

For conditions under which $\left( \tau_{LC} / \tau_{VAC} \sqrt{\mu} \right) \ll 1$, we would get:

$$R \simeq \rho n_g \sqrt{\sigma\eta} . \qquad (21)$$

On the other hand, if $\left( \tau_{LC} / \tau_{VAC} \sqrt{\mu} \right) >> 1$, we would get:

$$R \simeq \rho n_g \sqrt{\sigma\eta} \frac{\sqrt{\mu}}{\beta\sqrt{2}} \frac{\tau_{VAC}}{\tau_{LC}} . \qquad (22)$$

This limit corresponds to a case where the laser cavity finesse, $F$, is high. It is instructive in this case to consider a scenario where the length of the laser cavity is the same as the vacuum length difference between the two arms of the SLAUMZI. We also consider the ideal case of $\rho = 1, \beta = 1, \eta = 1, \sigma = 1, \mu = 1$. Under this condition, we see that:

$$R \simeq \frac{n_g}{\sqrt{2F}}. \tag{23}$$

This condition would be relevant if we require that the dimensions of the SLAUMZI are comparable to those of the laser cavity. Thus, under such a condition, the value of the group index need to exceed the finesse of the laser cavity significantly in order for the SLAUMZI technique to be useful in practice.

**Section B. Proposed SLAUMZI-based Gyroscope and Accelerometer**

Even though the application of the SLAUMZI can be broad, here we give a concrete example of how the SLAUMZI can be used for rotation and acceleration sensing. Figure - S1 shows a configuration for a SLAUMZI enhanced gyroscope and accelerometer. The left panel shows a pair of bi-directional Raman lasers, which can be either subluminal [5] or superluminal [6], employing gain from two separate cells, without any cross-talks. In order to take advantage of the SLAUMZI technique, it is necessary to employ a slow-light medium with a very large group index. One approach for achieving a group index as large as ten million is to make use of the process of electromagnetically induced transparency (EIT) in a Λ system [7]. For such a process, both legs of the beams have to be resonant with respect to the intermediate state. On the other hand, a Raman laser is always detuned by nearly a GHz away from any excited state. As such, in order to realize the EIT process, it is necessary to shift the Raman laser output with an acousto-optic modulator (AOM), for example, so that the shifted beam becomes resonant with one leg of the Λ system. In addition, it is necessary to ensure that the optical fields are phase coherent, even though they would differ in frequency by the ground state splitting. To fulfill this requirement, it is necessary to make use of another laser for the second leg, and off-set phase lock it to the Raman laser.

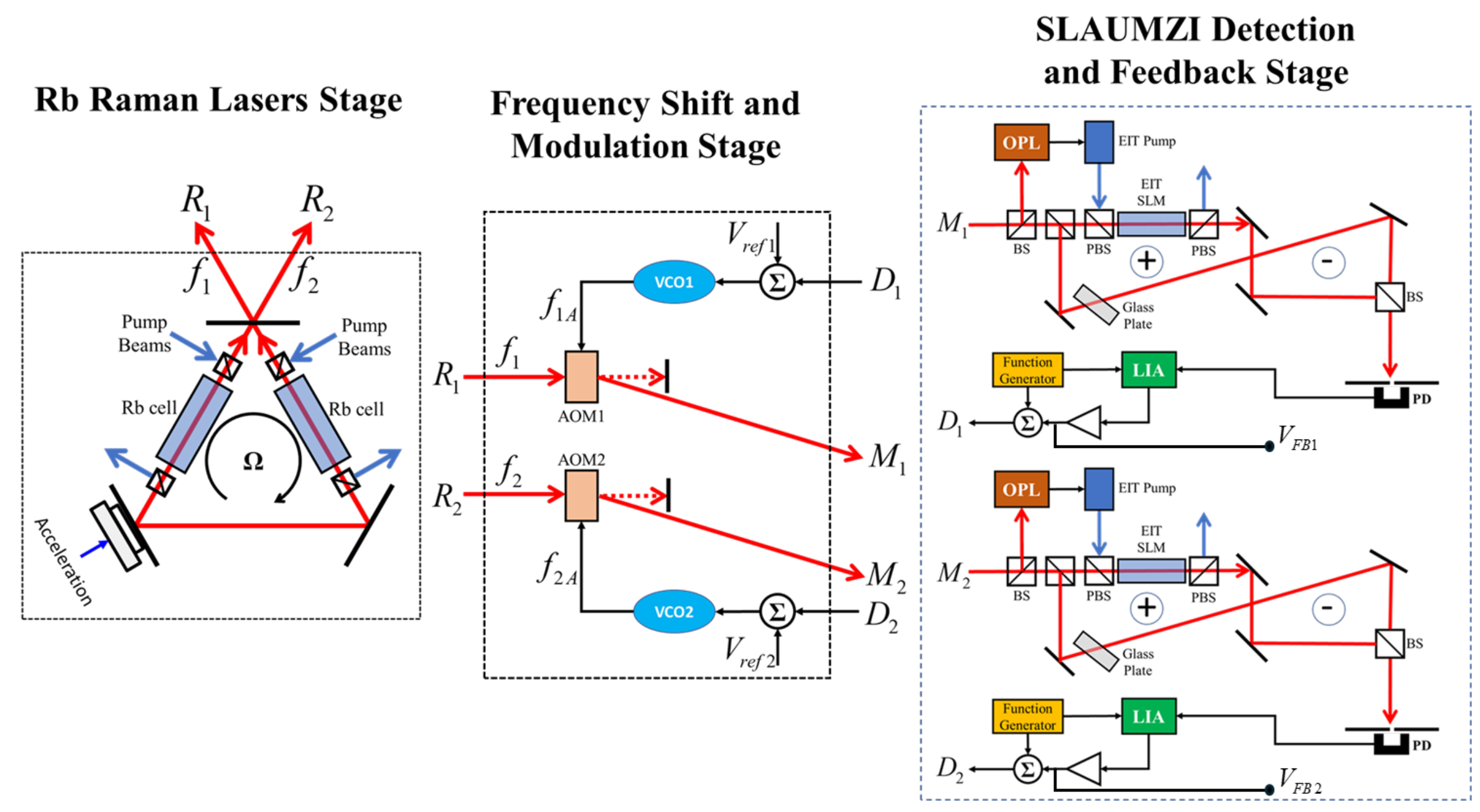

Figure - S1: The schematic of an accelerometer-incorporated slow-light enhanced gyroscope used for rotation sensing and accelerometry. The details are explained in the text. Note that each MZI is configured in the "figure 8" shape to eliminate the Sagnac effect.

There are two laser outputs $R_1$ and $R_2$ through the output coupler. Assume the lasing frequencies of these two laser outputs are $f_1$ and $f_2$, respectively. Consider first one of the outputs, for example, $R_1$. This is passed through an acoustic optical modulator (AOM) to produce $M_1$ with a frequency shift, which is denoted as $f_{1A}$ and is determined by the voltage reference $V_{ref1}$ in the voltage controlled oscillator (VCO). The $M_1$ beam will have the frequency at which the EIT effect can be produced. Before being guided to MZI, a pump laser is phase locked to $M_1$ using an off-set phase lock (OPL), where the offset is the hyperfine splitting between two ground states in $^{85}$Rb atoms, i.e. 3.0357GHz. To keep the MZI fringe at its peak intensity, a square-wave modulation is applied to the AOM, along with $V_{ref1}$ and the output of the servo after the lock-in amplifier (LIA), denoted as $V_{FB1}$. For the other output $R_2$, the same procedure is followed, using another SLAUMZI system. Without any rotation, the servo outputs for both $R_1$ and $R_2$ are zeroes (mean values). However, in the presence of rotation (around the axis normal to the area of the resonator), the frequencies of the output lasers would deviate from the original values with opposite signs due to the Sagnac effect, which would make $V_{FB1}$ and $V_{FB2}$ non-zero. For the case where there is only acceleration (in the direction perpendicular to the diaphragm mounted mirror), the frequency shift in each laser output would be the same. As such, the rate of rotation will be proportional to ($V_{FB1}$ - $V_{FB2}$), while the rate of acceleration would be proportional to ($V_{FB1}$ + $V_{FB2}$)/2.

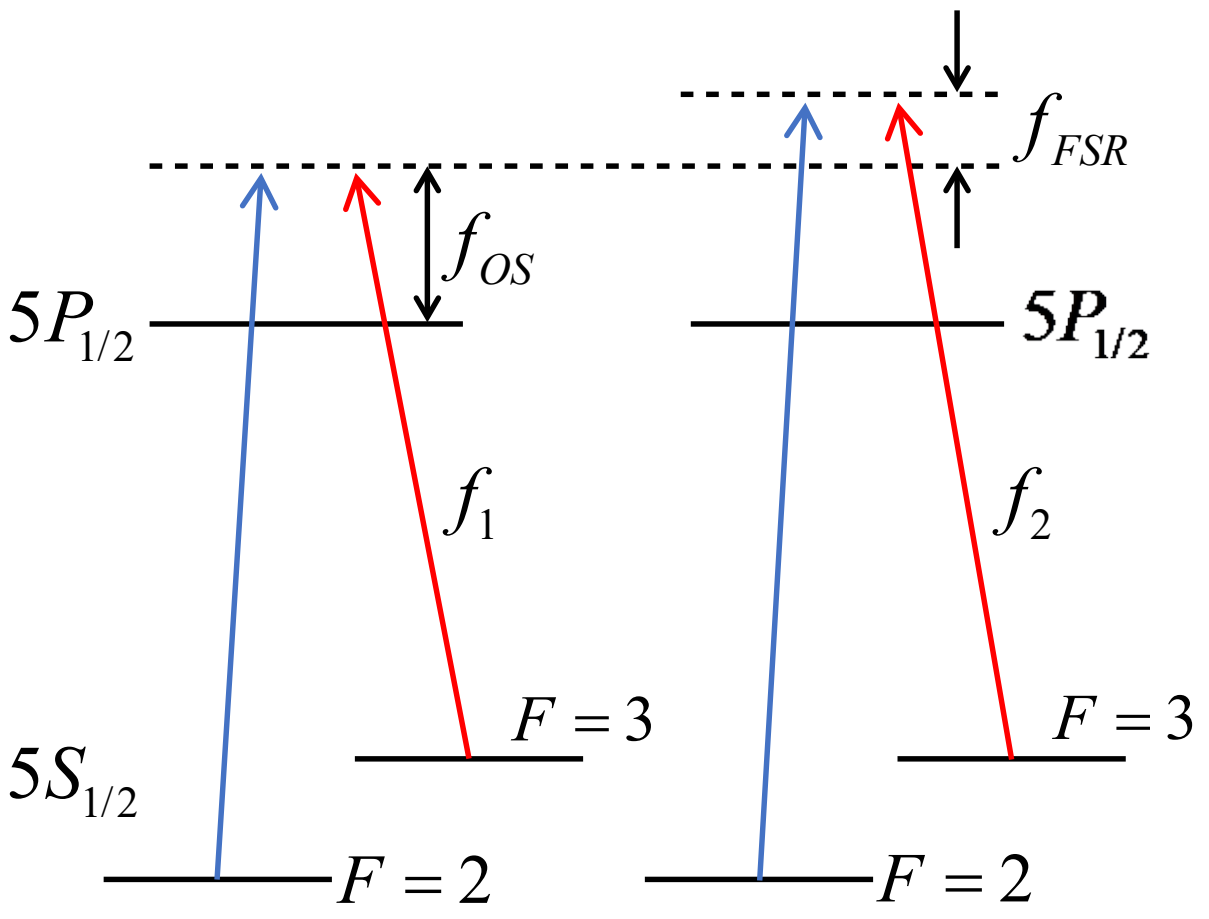

Figure - S2: Schematic of relevant energy levels and the laser fields for the ring laser stage. The optical pumps are not shown in the figure, but they create population inversion between the two ground states by coupling the $5S_{1/2}$, F= 3 to $5P_{3/2}$ transition.

In the design of such a device, the frequencies of the two Raman lasers can be non-degenerate, differing in frequency by one free spectral range (FSR) of the cavity. This will eliminate the so-called lock-in effect that occurs in a ring laser gyroscope [8], which keeps the two frequencies identical until a sufficiently large rotation occurs. Cumbersome techniques such as mechanical dithering are used to eliminate this effect in a He-Ne ring laser gyroscope. Use of the non-degeneracy can eliminate the need for such techniques. The non-degeneracy of the two Raman lasers is illustrated schematically in Figure - S2, which corresponds to the left panel in Figure - S1.

It should be noted that the Raman laser behaves as a subluminal laser [5], which suppresses the sensitivity by the factor of the group index inside the Raman laser. However, according to the theoretical model developed by Henry [9], the STL is reduced by a factor of the square of the group index. For a conventional measurement process, we recall that the MMFS is proportional to the square root of the STL. Thus, the MMFS decreases by a factor of the group index. As such, if Henry's model holds true, then the net frequency sensitivity of Raman laser under conventional detection will be unaffected by the group index inside the Raman laser. An extension of our analysis for the MMFS for the SLAUMZI shows that the group index inside the Raman laser has no effect on our result either, assuming Henry's model holds. Alternatively, the Raman lasers can be designed to operate superluminally, using any of the various approaches we have demonstrated so far [10,11,12,13,14,15]. In that case, the measurement sensitivity would be further enhanced by the superluminal laser enhancement factor [6].

---

[1] S. M. Shahriar, R. Zhu, J. Li and Z. Zhou, "The General Quantum Limit for and the Optimization of the Minimum Measurable Frequency Shift in a Laser," https://arxiv.org/pdf/2607.10518

[2] M. O. Scully and M.S. Zubairy, Quantum Optics, Cambridge University Press (1997).

[3] J. Capmany, J. Mart and H. Mangraham, "Impact of Finite Laser Linewidth on the Performance of OFDM Networks Employing Single-Cavity Fabry-Perot Demultiplexers," Journal of Lightwave Technology, vol. 13, no. 2 (1995).

[4] W. M. Itano, J. C. Bergquist, J. J. Bollinger, J. M. Gilligan, D. J. Heinzen, F. L. Moore, and D. J. Wineland, "Quantum projection noise: Population fluctuations in two-level systems," Phys. Rev. A 47, 3554 (1993).

[5] J. Scheuer and S.M. Shahriar, "Lasing dynamics of super and sub luminal lasers", Opt. Express 23, 32350– 32365 (2015).

[6] Z. Zhou, R. Zhu, N. Condon, D. Hileman, J. Bonacum and S.M. Shahriar, "Bi-directional Superluminal Ring Lasers without Cross-talk and Gain Competition," Appl. Phys. Lett. 120, 251105 (2022).

[7] L. V. Hau, S. E. Harris, Z. Dutton and C. H. Behrooz, "Light speed reduction to 17 metres per second in an ultracold atomic gas," Nature 397, 18 (1999).

[8] W. W. Chow, J. Gea-Banacloche, L. M. Pedrotti, V. E. Sanders, W. Schleich, and M. O. Scully, "The ring laser gyro," Reviews of Modern Physics 57, 61 (1987).

[9] C. H. Henry, "Theory of the Linewidth of Semiconductor Lasers," IEEE J. Of Quantum Electronics, Vol. QE-18, No. 2, (1982).

[10] Z. Zhou, M. Zhou and S.M. Shahriar, "A superluminal Raman laser with enhanced cavity length sensitivity," Optics Express 27, 29739-29745 (2019).

[11] Y. Sternfeld, Z. Zhou, J. Scheuer, and M. S. Shahriar, "Electromagnetically induced transparency in Raman gain for realizing a superluminal ring laser," Opt. Express 29, 1125-1139 (2021).

[12] Zifan Zhou, Nicholas Condon, Devin Hileman, and M. S. Shahriar, "Observation of a highly superluminal laser employing optically pumped Raman gain and depletion," Opt. Express 30, 6746-6754 (2022).

[13] Z. Zhou, R. Zhu, N. Condon, D. Hileman, J. Bonacum and S.M. Shahriar, "Bi-directional Superluminal Ring Lasers without Cross-talk and Gain Competition," Appl. Phys. Lett. 120, 251105 (2022).

[14] Z. Zhou, R. Zhu, Y. Sternfeld, J. Scheuer, J. Bonacum, and S. M. Shahriar, "Demonstration of a Superluminal Laser using Electromagnetically Induced Transparency in Raman Gain," Opt. Express 31(9), 14377-14388 (2023).

[15] Y. Sternfeld, Z. Zhou, S.M. Shahriar and J. Scheuer, "Single-pumped gain profile for a superluminal ring laser," Optics Express 31(22) 36952-36965 (2023).